\documentclass[pdflatex,sn-mathphys-num]{sn-jnl}

\usepackage{graphicx}%
\usepackage{multirow}%
\usepackage{amsmath,amssymb,amsfonts}%
\usepackage{amsthm}%
\usepackage{mathrsfs}%
\usepackage[title]{appendix}%
\usepackage{xcolor}%
\usepackage{textcomp}%
\usepackage{manyfoot}%
\usepackage{booktabs}%
\usepackage{algorithm}%
\usepackage{algorithmicx}%
\usepackage{algpseudocode}%
\usepackage{listings}%

\theoremstyle{thmstyleone}%
\theoremstyle{thmstyletwo}%

\theoremstyle{thmstylethree}%

\begin{document}

\title[Article Title]{Vision Foundation Models in Medical Image Analysis: Advances and Challenges}


\author[1]{\fnm{Pengchen} \sur{Liang}}\email{liangpengchen@shu.edu.cn}
\equalcont{These authors contributed equally to this work.}

\author*[2]{\fnm{Bin} \sur{Pu}}\email{eebinpu@ust.hk}

\author[3]{\fnm{Haishan} \sur{Huang}}\email{huanghsh25@mail2.sysu.edu.cn}
\equalcont{These authors contributed equally to this work.}

\author[4]{\fnm{Yiwei} \sur{Li}}\email{liyiwei@shchildren.com.cn}
\equalcont{These authors contributed equally to this work.}

\author[2]{\fnm{Hualiang} \sur{Wang}}\email{hwangfd@connect.ust.hk}

\author[5]{\fnm{Weibo} \sur{Ma}}\email{mwb1030@sina.com}

\author*[6]{\fnm{Qing} \sur{Chang}}\email{robie0510@hotmail.com}

\affil[1]{\orgdiv{School of Microelectronics}, \orgname{Shanghai University}, \orgaddress{\city{Shanghai}, \postcode{201800}, \country{China}}}

\affil[2]{\orgdiv{Department of Electronic and Computer Engineering}, \orgname{The Hong Kong University of Science and Technology}, \orgaddress{\city{Hong Kong}, \state{SAR}, \country{China}}}

\affil[3]{\orgdiv{School of Software Engineering}, \orgname{Sun Yat-sen University}, \orgaddress{\city{Zhuhai}, \postcode{519000}, \country{China}}}

\affil[4]{\orgdiv{Department of Nuclear Medicine, Shanghai Children's Hospital, School of Medicine}, \orgname{Shanghai Jiao Tong University}, \orgaddress{\city{Shanghai}, \postcode{200062}, \country{China}}}

\affil[5]{\orgdiv{School of Public Administration}, \orgname{East China Normal University}, \orgaddress{\city{Shanghai}, \postcode{200062}, \country{China}}}

\affil[6]{\orgdiv{Department Shanghai Key Laboratory of Gastric Neoplasms, Ruijin Hospital}, \orgname{Shanghai Jiao Tong University School of Medicine}, \orgaddress{\city{Shanghai}, \postcode{200025}, \country{China}}}



\abstract{The rapid development of Vision Foundation Models (VFMs), particularly Vision Transformers (ViT) and Segment Anything Model (SAM), has sparked significant advances in the field of medical image analysis. These models have demonstrated exceptional capabilities in capturing long-range dependencies and achieving high generalization in segmentation tasks. However, adapting these large models to medical image analysis presents several challenges, including domain differences between medical and natural images, the need for efficient model adaptation strategies, and the limitations of small-scale medical datasets. This paper reviews the state-of-the-art research on the adaptation of VFMs to medical image segmentation, focusing on the challenges of domain adaptation, model compression, and federated learning. We discuss the latest developments in adapter-based improvements, knowledge distillation techniques, and multi-scale contextual feature modeling, and propose future directions to overcome these bottlenecks. Our analysis highlights the potential of VFMs, along with emerging methodologies such as federated learning and model compression, to revolutionize medical image analysis and enhance clinical applications. The goal of this work is to provide a comprehensive overview of current approaches and suggest key areas for future research that can drive the next wave of innovation in medical image segmentation.}




\maketitle

\section{Introduction}

In recent years, the application of Vision Foundation Models (VFMs) in medical image analysis has witnessed remarkable progress, especially with the advent of Vision Transformers (ViT) and Segment Anything Model (SAM) \cite{shamshad2023transformers,wu2023medical,kirillov2023segment}. 
These models have shown exceptional performance in capturing long-range dependencies, which has been a challenge for traditional Convolutional Neural Networks (CNNs) due to their inherent limitations in modeling spatial relationships \cite{wang2022mixed,shamshad2023transformers,liang2024data,pu2024m3}. 
The introduction of Transformer-based models, such as TransUNet and Swin-UNet, has significantly enhanced the performance of medical image segmentation tasks by combining global attention mechanisms with the precise localization abilities of U-Net architectures \cite{chen2021transunet,cao2022swin}. 
However, despite their impressive capabilities, adapting these models to medical contexts presents several challenges, especially due to the inherent differences between medical and natural images.

One of the key challenges in applying VFMs to medical image segmentation is domain adaptation \cite{wu2023medical}. 
The large-scale datasets required for pretraining these models are often not available in the medical field due to the high cost and time constraints of acquiring labeled medical images \cite{zhang2024efficientvit}. As a result, fine-tuning these models on smaller medical datasets often leads to performance degradation due to domain mismatch \cite{wu2023segment,huang2025learnable}. 
To address this, researchers have proposed various strategies, including the use of adapter modules and cross-domain transfer learning to improve the adaptability of VFMs to medical images \cite{li2025stitching,hu2025spa}.

Another challenge is the need for computationally efficient models that can be deployed on edge devices in clinical settings. Given the resource constraints of medical edge devices, techniques such as model compression and knowledge distillation are becoming increasingly important \cite{zhang2024efficientvit,shi2024knowledge}. Knowledge distillation, in particular, has emerged as a promising approach to transferring the capabilities of large, pre-trained models to smaller, more efficient models without sacrificing performance \cite{shi2024knowledge}. 
This has become a key area of research as models like SAM, CLIP, and others continue to evolve.

In addition to these challenges, the integration of VFMs with Federated Learning (FL) has opened up new possibilities for collaborative training of models across distributed medical institutions while preserving patient privacy  \cite{wu2023fedms}. Federated Learning provides an opportunity to overcome the issue of limited data availability and privacy concerns by enabling models to learn from decentralized data without sharing raw patient data \cite{zhuang2023foundation}. 
This is particularly important in the medical field, where data privacy is a critical concern.

This paper reviews state-of-the-art research on the adaptation of VFMs in medical image analysis, focusing on the challenges and solutions related to domain adaptation, federated learning, model compression, and knowledge distillation. We also propose future research directions aimed at overcoming these challenges and advancing the application of VFMs in medical image analysis.

\begin{figure*}[htbp]
\centering
\includegraphics[width=\textwidth]{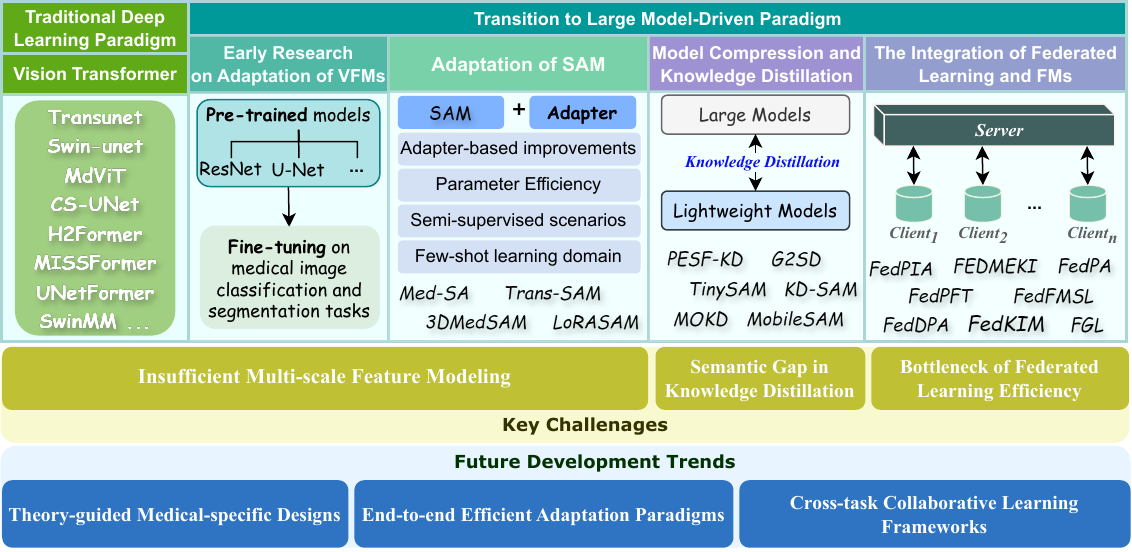}
\caption{An overview of vision foundation models (VFMs) in medical image analysis.   The top panel showcases the technical framework of VFMs in medical image analysis, focusing on advances related to domain adaptation, federated learning, model compression, and knowledge distillation.   The middle panel highlights three key challenges.  The bottom panel outlines the future development trends.}
\label{fig:fig1}
\end{figure*}

\section{Vision Transformer in Medical Image Analysis}

In recent years, Vision Transformer (ViT) has made significant progress in the field of medical image analysis due to its excellent modeling capabilities and its ability to capture long-range dependencies \cite{shamshad2023transformers,liang2024rskd,liang2023dawtran}. 

Traditional Convolutional Neural Networks (CNNs), although highly successful in medical image segmentation tasks, have limitations due to the inherent restrictions of convolution operations, which struggle to capture long-range dependencies \cite{wang2022mixed,pu2021fetal,pu2022mobileunet,zhao2023transfsm,pu2024hfsccd}. 
To address this issue, researchers have proposed various innovative Transformer-based architectures. Among them, TransUNet has become a landmark work by combining the global self-attention mechanism of Transformer with the precise localization ability of U-Net, demonstrating outstanding performance in tasks such as multi-organ segmentation \cite{chen2021transunet}. 
Subsequently, Swin-UNet further improved the model’s efficiency and performance by introducing a hierarchical Transformer structure and a shifted window mechanism \cite{cao2022swin}. In response to the challenge of small-scale medical image datasets, several improvements have been proposed: MDViT improves the model’s performance on small datasets by employing a multi-domain learning strategy \cite{du2023mdvit}; CS-UNet enhances the spatial modeling ability of Transformer by incorporating convolution operations \cite{liu2023optimizing}. Additionally, to balance computational efficiency and segmentation accuracy, researchers have developed various hybrid architectures: H2Former combines the local feature extraction capabilities of CNNs with the global modeling capabilities of Transformer, achieving excellent segmentation performance while maintaining low computational complexity \cite{he2023h2former}; MISSFormer redefines Transformer blocks and feature fusion strategies, achieving significant breakthroughs in multi-organ segmentation tasks \cite{huang2022missformer}. In 3D medical image segmentation, models such as UNetFormer and SwinMM have effectively modeled 3D spatial information through innovative architectural designs \cite{hatamizadeh2022unetformer, wang2023swinmm}. 

These works demonstrate that Transformer has great potential in medical image analysis. Through reasonable architectural design and optimization strategies, it can effectively overcome challenges such as data scale limitations and computational efficiency, providing better solutions for medical image analysis tasks. Despite the significant progress of ViT in the medical field, such models often require pretraining from scratch, and due to the high cost of annotating medical data, it is challenging to fully leverage their representational potential.

\section{Transition to Large Model-Driven Paradigm in Medical Image Analysis}

With the breakthrough progress of Vision Foundation Models (VFMs) in computer vision, the medical image analysis field is undergoing a profound transformation from traditional deep learning paradigms to large model-driven paradigms \cite{mazurowski2023segment,lee2024foundation,moor2023foundation,khan2025comprehensive,waqas2023revolutionizing,rood2024human,wang2023clipn,wang2024tri}. As a core technological representative of this transition, the Segment Anything Model (SAM), with its strong zero-shot segmentation ability and universal representation characteristics, has gradually demonstrated unique advantages in medical image processing \cite{wu2023medical,ma2024segment,gao2024desam}. However, due to the special nature of medical images (such as multi-modal imaging, complex anatomical structures, data privacy constraints, etc.), achieving effective adaptation of SAM and other foundational models in medical contexts still faces many challenges \cite{feng2025swinsam,zhou2025segment}. 

This section systematically reviews the research progress in adapting foundational models to medical contexts, including the technical paths for model adaptation, model compression, and optimization under federated learning frameworks, revealing the current technological bottlenecks and providing theoretical support for the innovative direction of this research.

\subsection{Early Research on Adaptation of Vision Foundation Models in Medical Image Analysis}

Research on the adaptation of vision foundation models to medical image analysis began with the transfer of pre-trained natural image models \cite{morid2021scoping,wang2023pre}. Early work mainly focused on fine-tuning ImageNet pre-trained models, such as ResNet and U-Net, to improve their adaptability in medical image classification and segmentation tasks \cite{morid2021scoping}. For example, Kalinin et al. \cite{kalinin2020medical} significantly improved the performance of the U-Net model in tasks such as abnormal vascular development segmentation in wireless capsule endoscopy videos and semantic segmentation of surgical instruments in robotic surgery videos by incorporating an ImageNet pre-trained encoder, demonstrating the effectiveness of pre-training strategies in medical image segmentation. However, such methods generally encounter feature representation bias issues when facing the significant domain differences between medical images and natural images, leading to a decline in performance after fine-tuning \cite{guan2021domain,davila2024comparison}.

\subsection{The Emergence of SAM and Its Challenges in Medical Image Segmentation}

The release of SAM, based primarily on ViT, marks the entry of vision foundation models into a new era of "universal segmentation." Its segmentation engine, trained on 11 million natural images, exhibits remarkable generalization ability in zero-shot scenarios \cite{kirillov2023segment}. 

However, due to the significant domain differences between medical and natural images, directly applying SAM often does not yield ideal results. To address this, researchers have proposed several adapter-based improvements \cite{lei2025medlsam,fan2025ma,huang2025learnable,li2025stitching,hu2025spa}. For example, Wu et al. \cite{wu2023medical} first proposed the Medical SAM Adapter (Med-SA), achieving efficient domain adaptation through Space-Depth Transpose and Hyper-Prompting Adapter. To address the specificity of 3D medical images, Lin et al. \cite{lin2025volumetric} introduced the 3D Medical SAM-Adapter (3DMedSAM), which innovatively designs a 3D patch embedding module and a multi-scale 3D mask decoder to achieve cross-dimensional adaptation from 2D to 3D. 

In terms of parameter efficiency, Wu et al. \cite{wu2025trans} proposed Trans-SAM, which employs a Parameter-Efficient Fine-Tuning (PEFT) strategy, effectively integrating pre-trained features through Intuitive Perceptual Fine-tuning adapters and Multi-scale Domain Transfer adapters. Paranjape et al. \cite{paranjape2025low} proposed LoRASAM, using low-rank adaptation to reduce training parameters by over 99\%, significantly improving performance. 

Specific medical tasks have also seen improvements, such as Gu et al. \cite{gu2024lesam}, who proposed LeSAM for lesion segmentation, incorporating an improved mask decoder to achieve more precise boundary delineation. Shi et al. \cite{shi2024mask} designed Mask-Enhanced SAM (M-SAM) for tumor lesion segmentation by enriching medical image semantics through the Mask-Enhanced Adapter. Chen et al. \cite{chen2024ba} introduced BA-SAM, which incorporates a Boundary-Aware Attention module to significantly improve boundary recognition. 

For semi-supervised scenarios, Huang et al. \cite{huang2025learnable} proposed KnowSAM, which achieves more robust segmentation through Multi-view Co-training and Learnable Prompt Strategy. Lu et al. \cite{lu2024up} proposed UP-SAM, which innovatively considers both cognitive uncertainty and incidental uncertainty. 

In the few-shot learning domain, Xie et al. \cite{xie2024sam} proposed an improved strategy based on few-shot embedding, significantly reducing the annotation requirements. To enhance SAM's generalization ability, Gao et al. \cite{gao2024desam} proposed DeSAM, which alleviates the negative impact of poor prompts on mask generation through decoupling design. Li et al. \cite{li2025stitching} proposed the SFR framework, employing a three-stage strategy of stitching, fine-tuning, and retraining to achieve better 3D segmentation results. Notably, \cite{lyu2024mcp} recently proposed MCP-MedSAM, which lowers training resource requirements to the level of a single GPU day while maintaining competitive performance. 

In clinical applications, \cite{zhang2023segment} systematically evaluated SAM's performance in radiotherapy, validating its segmentation effect on different anatomical sites. Additionally, works by \cite{lin2023samus}, \cite{hu2025spa}, and \cite{wang2024samihs} have made significant progress in ultrasound image segmentation, spatial feature extraction, and intracranial hemorrhage segmentation, respectively, further confirming the broad application prospects of adapter-based SAM improvement methods in medical image segmentation. 

Despite the progress, existing adaptation methods still have three key limitations: (1) insufficient modeling of multi-scale contextual relationships in medical images, limiting the segmentation accuracy of small anatomical structures; (2) the lack of targeted parameter update strategies for domain features, which may lead to overfitting or under-adaptation; (3) many architectural improvements are based on heuristic designs, lacking systematic optimization guided by theory. These bottlenecks need to be overcome through innovations in foundational model adaptation theory.

\subsection{Model Compression and Knowledge Distillation in Medical Edge Devices}

The computational resource constraints of medical edge devices have led to a growing need for model compression techniques. Knowledge Distillation (KD), as a mainstream compression paradigm, transfers knowledge from large models (teachers) to smaller models (students), reducing inference costs while maintaining performance \cite{xuan2023distilling,xu2024survey,rao2023parameter}. 

In recent years, with the rapid development of vision foundation models such as SAM and CLIP, how to transfer the capabilities of these large models to lightweight models through knowledge distillation has become a hot research topic \cite{shu2023tinysam,zhang2023faster,song2024sam}. Xuan et al. \cite{xuan2023distilling} proposed a data-independent knowledge distillation method, synthesizing alternative data through diverse prompts. Shakir et al. \cite{shakir2024efficacy} explored the effectiveness of knowledge distillation based on foundational models in image classification tasks, finding that using the logits or feature representations of teacher models can significantly improve the performance of student models. Rao et al. \cite{rao2023parameter} proposed a parameter-efficient knowledge distillation method, PESF-KD, which adaptively adjusts the soft labels of teacher networks to achieve efficient knowledge transfer. In the self-supervised learning domain, Song et al. \cite{song2023multi} proposed a multi-mode online knowledge distillation method, MOKD, which achieves collaborative learning through self-distillation and cross-distillation modes. Huang et al. \cite{huang2023generic} proposed a two-stage distillation strategy, G2SD, for lightweight ViT models, ensuring task-specific performance while maintaining generalization. 

In the medical image domain, Shi et al. \cite{shi2024knowledge} distilled the knowledge of SAM into the U-Net model for medical image segmentation. Patil et al. \cite{patil2025efficient} proposed the KD-SAM framework, which jointly optimizes the encoder and decoder through a combination of MSE and perceptual loss. Wu et al. \cite{wu2023segment} demonstrated that SAM serves as a good teacher for local feature learning and proposed an auxiliary task using attention-weighted semantic relation distillation. Wang et al. \cite{wang2024exploring} explored the semantic prompting role of SAM in domain adaptation. To improve SAM's inference efficiency, researchers have proposed various lightweight solutions: \cite{zhang2023faster} proposed MobileSAM, replacing the heavy image encoder with a lightweight version through decoupled distillation; \cite{shu2023tinysam} proposed TinySAM, which uses full-stage knowledge distillation and quantization strategies; \cite{song2024sam} designed SAM-Lightening based on sparse flash attention, achieving a 30x speedup; \cite{zhang2024efficientvit} proposed EfficientViT-SAM, achieving a 48.9x speedup without sacrificing performance; \cite{liu2024pq} proposed the first post-training quantization method for SAM, PQ-SAM. 

Additionally, numerous innovative works have emerged for specific applications: \cite{chen2023make} proposed an unannotated shadow detection framework, ShadowSAM; \cite{zhang2024freekd} achieved knowledge distillation through semantic frequency prompting; \cite{xu2024multidimensional} explored the multidimensional applications of SAM in weakly supervised video saliency object detection. These studies show that knowledge distillation of foundational models is evolving toward more efficient and specialized directions, providing important support for the efficient deployment of foundational models in edge devices and specific scenarios.

\subsection{The Integration of Federated Learning and Foundation Models}

With the rapid development of foundational models (FMs), the integration of FMs with Federated Learning (FL) has become an important research direction in artificial intelligence. While foundational models have demonstrated outstanding performance in natural language processing, computer vision, and multimodal tasks \cite{wu2023fedms,guo2023promptfl}, their large parameter sizes and massive data requirements also pose significant challenges. Federated Learning, as a distributed training paradigm that protects data privacy, offers a potential solution to the data acquisition and privacy protection challenges faced by foundational models in practical applications \cite{ren2024advances,zhuang2023foundation}. 

In the medical field, the integration of foundational models and federated learning has shown tremendous potential. Research has shown that federated learning frameworks with foundational models have achieved significant results in multiple medical tasks, including cardiac CT image analysis \cite{tolle2024federated}, endoscopic surgery \cite{do2025fedefm}, ultrasound imaging \cite{jiang2024privacy}, and retinal age prediction \cite{nielsen2024foundation}. These applications not only improve diagnostic accuracy but also effectively address the issue of limited medical data sharing \cite{dayan2021federated}. To tackle the specific challenges of the medical field, researchers have proposed several innovative solutions, such as the FedKIM framework \cite{wang2024fedkim} and the FEDMEKI platform \cite{wang2024fedmeki}, which effectively handle multi-modal and heterogeneous medical data. 

At the technical level, several innovative methods have been proposed to optimize the performance of foundational models in federated learning. The introduction of Parameter-Efficient Fine-Tuning (PEFT) techniques significantly reduces communication overhead and computational burden \cite{sun2024exploring}. Methods such as FedPFT \cite{beitollahi2024parametric} and FedPIA \cite{saha2024fedpia} use innovative parameter sharing and integration strategies to drastically reduce resource consumption while maintaining model performance. In addition, the sparse activation LoRA algorithm proposed by FedFMSL \cite{wu2024fedfmsl} only requires adjustments to less than 0.3\% of the model parameters, achieving excellent performance. 

To address data heterogeneity, researchers have proposed dual personalization adapter architectures \cite{yang2024dual} and prompt-based federated learning methods \cite{zhang2023federated}. These methods effectively handle data distribution differences between clients and achieve better model personalization. 

In the recommendation system field, federated adaptation mechanisms have been designed to enhance the performance of foundational models \cite{zhang2024federated}. Future research will focus on improving communication efficiency, enhancing model robustness, protecting data privacy, and handling heterogeneous data \cite{woisetschlager2024survey,li2024synergizing}. Solving these challenges will further promote the deployment and development of federated foundational models in practical applications.

\section{Conclusion}

In conclusion, the adaptation of vision foundation models in medical image analysis has made initial progress, but several key challenges remain: (1) insufficient multi-scale feature modeling, which limits the segmentation accuracy of small structures in medical images; (2) the semantic gap in knowledge distillation, where domain differences between natural and medical images lead to distortion in knowledge transfer; (3) the bottleneck of federated learning efficiency, where traditional parameter compression strategies struggle to balance communication overhead with heterogeneous data adaptability.

The future development trends are characterized by three prominent features: (1) foundational model architecture innovation will shift from simple fine-tuning to theory-guided medical-specific designs; (2) privacy-preserving computation technologies will be deeply coupled with model compression, forming end-to-end efficient adaptation paradigms; (3) cross-task collaborative learning frameworks will break through the limitations of traditional single-task optimization, achieving joint enhancement in segmentation, restoration, and diagnosis.

\bibliography{sn-bibliography}

\end{document}